\documentclass[aip,jcp,superscriptaddress,amsmath,amssymb,twocolumn,reprint]{revtex4-1}

\usepackage[version=3]{mhchem} % Formula subscripts using \ce{}
\usepackage{graphicx}
\usepackage{amsmath}
\usepackage{amssymb}
\usepackage{nicefrac}
\usepackage{booktabs}
\usepackage{mwe}
\usepackage{array}
\newcolumntype{m}{>{$} c <{$}}
\usepackage{color}

\def\rv{{\bf r}}

\def\beq{\begin{equation}}
\def\eeq{\end{equation}}

\begin{document}     

\author{Michael Seidl}
\affiliation
{Department of Theoretical Chemistry and Amsterdam Center for Multiscale Modeling, Faculty of Science, Vrije Universiteit, De Boelelaan 1083, 1081HV Amsterdam, The Netherlands}
\author{Sara Giarrusso}
\affiliation
{Department of Theoretical Chemistry and Amsterdam Center for Multiscale Modeling, Faculty of Science, Vrije Universiteit, De Boelelaan 1083, 1081HV Amsterdam, The Netherlands}
\author{Stefan Vuckovic}
\affiliation
{Department of Theoretical Chemistry and Amsterdam Center for Multiscale Modeling, Faculty of Science, Vrije Universiteit, De Boelelaan 1083, 1081HV Amsterdam, The Netherlands}
\affiliation{Department of Chemistry, University of California, Irvine, CA 92697, USA}
\author{Eduardo Fabiano}
\affiliation{Institute for Microelectronics and Microsystems (CNR-IMM), Via Monteroni, Campus Unisalento, 73100 Lecce, Italy}
\affiliation{Center for Biomolecular Nanotechnologies @UNILE, Istituto Italiano di Tecnologia, Via Barsanti, I-73010
Arnesano, Italy}
\author{Paola Gori-Giorgi}
\affiliation
{Department of Theoretical Chemistry and Amsterdam Center for Multiscale Modeling, Faculty of Science, Vrije Universiteit, De Boelelaan 1083, 1081HV Amsterdam, The Netherlands}
%\email{}

\title{Strong-interaction limit of an adiabatic connection in Hartree-Fock theory}
\begin{abstract}
We show that the leading term in the strong-interaction limit of the adiabatic connection that has as weak-interaction expansion the M\o ller-Plesset perturbation theory can be fully determined from a functional of the Hartree-Fock density. We analyze this functional and highlight similarities and differences with the strong-interaction limit of the density-fixed adiabatic connection case of Kohn-Sham density functional theory.
\end{abstract}

\maketitle

%%%%%%%%%%%%%%%%%%%%%%%%%%%%%%%%%%%%%%%%%%%%%%%%%%%%%%%%%%%%%%%%%%%%%
%% The manuscript does not need to include \maketitle, which is
%% executed automatically.  The document should begin with an
%% abstract, if appropriate.  If one is given and should not be, the
%% contents will be gobbled.
%%%%%%%%%%%%%%%%%%%%%%%%%%%%%%%%%%%%%%%%%%%%%%%%%%%%%%%%%%%%%%%%%%%%%
The adiabatic connection formalism has always been a powerful theoretical tool to build approximations for the exchange-correlation (XC) energy in density functional theory (DFT), in particular for hybrid\cite{Bec-JCP-93a,Bec-JCP-93,PerErnBur-JCP-96} and double-hybrid functionals,\cite{ShaTouSav-JCP-11,Gri-JCP-06,LarGri-JCTC-10,SuXu-JCP-14} but also for other kinds of XC functionals.\cite{Ern-CPL-96,SeiPerKur-PRL-00,MorCohYan-JCP-06,Bec-JCP-13a,locpaper,VucIroWagTeaGor-PCCP-17,BahZhoErn-JCP-16,VucGor-JPCL-17} In the Kohn-Sham (KS) framework, the density fixed adiabatic connection can be defined via the $\lambda$-dependent hamiltonian,\cite{LanPer-SSC-75,GunLun-PRB-76}
\begin{equation}\label{eq:adiabDFT}
	\hat{H}_{\lambda}^{\rm DFT}=\hat{T}+\lambda\,\hat{V}_{ee}+\hat{V}_{\lambda}[\rho],
\end{equation} 
where $\hat{T}$ is the kinetic energy operator for the $N$ electrons, $\hat{V}_{ee}$ is their mutual Coulomb repulsion, and $\hat{V}_{\lambda}[\rho]=\sum_{i=1}^N v_{\lambda}(\rv_i,[\rho])$ is the one body potential that makes the ground-state wavefunction of Eq.~\eqref{eq:adiabDFT}, $\Psi_{\lambda}^{\rm DFT}$, yield the density $\rho(\rv)\equiv\rho_{\lambda=1}(\rv)$ for all values of $\lambda$. From Eq.~\eqref{eq:adiabDFT}, one can derive an exact formula for the KS DFT XC energy,\cite{LanPer-SSC-75,GunLun-PRB-76}
\begin{equation}\label{eq:CCIDFT}
	E_{xc}^{\rm DFT}[\rho]=\int_0^1 W_{\lambda}^{\rm DFT}[\rho]\,d\lambda,
\end{equation}
where
\begin{equation}\label{eq:WlambdaDFT}
	W_{\lambda}^{\rm DFT}[\rho]\equiv\langle\Psi_{\lambda}^{\rm DFT}[\rho]|\hat{V}_{ee}|\Psi_{\lambda}^{\rm DFT}[\rho]\rangle - U[\rho],
\end{equation}
with $U[\rho]$ the Hartree energy. The coupling-constant integrand of Eq.~\eqref{eq:WlambdaDFT} has the known small\cite{GorLev-PRB-93} and large-$\lambda$ expansions \cite{GorVigSei-JCTC-09}
\begin{eqnarray}
W_{\lambda\to 0}^{\rm DFT}[\rho] & = & E_x^{\rm DFT} +\sum_{n=2}^\infty n\,E_c^{{\rm GL}n}\,\lambda^{n-1}, \label{eq:lambda0DFT} \\ 
\label{eq:lambdainfDFT} 
W_{\lambda\rightarrow\infty}^{\rm DFT}[\rho] & = & W_\infty^{\rm DFT}[\rho] + \frac{1}{\sqrt{\lambda}} W_{\infty}'^{\rm DFT}[\rho] + \cdots\ ,
\end{eqnarray}
where $E_x^{\rm DFT}$ is the exact KS exchange energy (the same expression as in Hartree-Fock theory, but using KS orbitals) and $E_c^{{\rm GL}n}$ is the $n^{\rm th}$ term in the G\"orling-Levy perturbation series.\cite{GorLev-PRB-93,GorLev-PRA-94} The expansion for large $\lambda$ of Eq.~\eqref{eq:lambdainfDFT} has as leading term the functional $W_\infty^{\rm DFT}[\rho]$, given by the minimum possible expectation value of the electron-electron repulsion in a given density $\rho(\rv)$,\cite{SeiGorSav-PRA-07,Lew-CRM-18,CotFriKlu-ARMA-18}
\begin{equation}
W_\infty^{\rm DFT}[\rho] = \inf_{\Psi\to\rho}\langle\Psi|\hat{V}_{ee}|\Psi\rangle-U[\rho],
\label{eq:WinftyDFT}
\end{equation}
while the next leading term, determined by $W_{\infty}'^{\rm DFT}[\rho]$, corresponds to the potential energy of zero-point vibrations around the manifold corresponding to the support of the minimizing probability density in Eq.~\eqref{eq:WinftyDFT}.\cite{GorVigSei-JCTC-09} While for the leading term there are rigorous proofs, \cite{Lew-CRM-18,CotFriKlu-ARMA-18} this next term is only a very reasonable conjecture that has been confirmed numerically in simple cases where it was possible to compute the exact integrand $W_{\lambda}^{\rm DFT}[\rho]$ .\cite{CorKarLanLee-PRA-17,GroKooGieSeiCohMorGor-JCTC-17}

Mixing KS DFT with Hartree-Fock (HF) ingredients is an approximation strategy that has a long history in chemistry, already starting with hybrids\cite{Bec-JCP-93a,Bec-JCP-93,PerErnBur-JCP-96,HeyScuErn-JCP-03,ZhaTru-ACR-08,JarScuErn-JCP-03,ArbKau-CPL-07} and double hybrids,\cite{Gri-JCP-06,LarGri-JCTC-10,ShaTouSav-JCP-11,SuXu-JCP-14} but also by simply inserting the HF density into a given approximate XC density functional.\cite{GillJohPopFri-IJQC-92,OliBar-JCP-94,KimSimBur-JCP-11,KimSimBur-JCP-14,KimSimBur-PRL-13,SimSonBur-JPCL-18} Very recently, it has also been observed that rather accurate interaction energies,\cite{FabGorSeiDel-JCTC-16,GiaGorDelFab-JCP-18} particularly for non-covalent complexes,\cite{VucGorDelFab-JPCL-18} can be obtained from models for $W_{\lambda}^{\rm DFT}[\rho]$ that interpolate between the two limits of Eq.~\eqref{eq:lambda0DFT} -- retaining only the first term, GL2,  in the GL series -- and of Eq.~\eqref{eq:lambdainfDFT}, using HF densities and orbitals as input, i.e., by constructing {\em de facto} an approximate resummation of the M\o ller-Plesset (MP) series, a procedure that lacks so far a theoretical justification. Motivated in particular by these last findings, we analyze in this communication the Hartree-Fock adiabatic connection [Eq.~\eqref{eq:HlambdaHF} below] whose Taylor expansion around $\lambda=0$ is the MP series [Eq.~\eqref{eq:lambda0HF} below] and show that the leading term in the $\lambda\to\infty$ expansion is determined by a functional of the HF density, see Eqs.~\eqref{eq:PsiHFlambdainf} and \eqref{eq:lambdainfHFall} below. We also highlight similarities and differences with the DFT case, showing that the large $\lambda$ expansion in HF theory has a structure similar to the one of Eq.~\eqref{eq:lambdainfDFT}.

We keep the notation general, as only few key properties of the HF operators are important here. We consider the adiabatic connection (see, e.g., Ref~\onlinecite{Per-IJQC-18})
\begin{equation}\label{eq:HlambdaHF}
	\hat{H}_{\lambda}^{\rm HF}=\hat{T}+\hat{V}_{\rm ext}+\hat{J}+\hat{K}+\lambda(\hat{V}_{ee}-\hat{J}-\hat{K}),
\end{equation}
with $\hat{V}_{\rm ext}$ the (nuclear) external potential and $\hat{J}=\hat{J}[\rho^{\rm HF}]$ and $\hat{K}=\hat{K}[\{\phi_i^{\rm HF}\}]$ the standard HF Coulomb and exchange operators, which are fixed once for all in the initial HF calculation, and do not depend on $\lambda$, but only on the HF density $\rho^{\rm HF}$ and occupied HF orbitals $\{\phi_i^{\rm HF}\}$. In the ground state $\Psi_{\lambda}^{\rm HF}$ of $\hat{H}_{\lambda}^{\rm HF}$, the density $\rho_\lambda(\rv)$ changes with $\lambda$: $\rho_{\lambda=0}(\rv)$ is the HF density $\rho^{\rm HF}(\rv)$, and $\rho_{\lambda=1}(\rv)$ is the exact physical density $\rho(\rv)$. Note that Teale {\em et al}.\cite{TeaCorHel-JCP-09} have analyzed a related adiabatic connection, in which the external potential is kept fixed; in that framework in the limit $\lambda\to\infty$ all the electrons but one escape to infinity. From Eq.~\eqref{eq:HlambdaHF}, the  Hellmann-Feynman theorem yields the exact formula
\begin{equation}
	E_{xc}^{\rm HF}=\int_0^1 W_{\lambda}^{\rm HF}\,d\lambda
\end{equation}
for the XC energy in the HF framework, with 
\begin{equation}\label{eq:WlambdaHF}
	W_{\lambda}^{\rm HF}\equiv\langle\Psi_{\lambda}^{\rm HF}|\hat{V}_{ee}-\hat{J}-\hat{K}|\Psi_{\lambda}^{\rm HF}\rangle + U[\rho^{\rm HF}]+2 E_x^{\rm HF}.
\end{equation}
Eq.~\eqref{eq:WlambdaHF} has been defined to allow for a direct comparison with the DFT $W_{\lambda}^{\rm DFT}[\rho]$ of Eqs.~\eqref{eq:CCIDFT} and \eqref{eq:WlambdaDFT}, with $W_{\lambda=0}^{\rm HF}=E_x^{\rm HF}$, and for small $\lambda$
\begin{equation}\label{eq:lambda0HF}
	W_{\lambda\to0}^{\rm HF}  =  E_x^{\rm HF} +\sum_{n=2}^\infty n\,E_c^{{\rm MP}n}\,\lambda^{n-1},
\end{equation}
where $E_c^{{\rm MP}n}$ the $n^{\rm th}$ term in the MP series. As is well-known (see, e.g., Refs.~\onlinecite{OlsChrKocJor-JCP-96,ForHeHeCre-IJQC-00}), the radius of convergence of the MP series is in general smaller than 1. Here we ask the question: {\em what happens to $\Psi_{\lambda}^{\rm HF}$ and $W_{\lambda}^{\rm HF}$ as $\lambda\to\infty$?} After answering this theoretical question, we will discuss its actual relevance for constructing approximations. 

When $\lambda$ becomes very large, the term $\lambda(\hat{V}_{ee}-\hat{J}-\hat{K})$ in Eq.~\eqref{eq:HlambdaHF} becomes more and more important, and we argue that the wavefunction $\Psi_{\lambda}^{\rm HF}$ should end up minimizing this term alone, similarly to the DFT case\cite{SeiGorSav-PRA-07} of Eq.~\eqref{eq:WinftyDFT}. The difference here is that the minimizer is not constrained to yield a fixed density, and the operator to be minimized also contains $-\hat{J}-\hat{K}$. We further argue that the expectation value of $\hat{K}$ is subleading with respect to the one of $\hat{V}_{ee}-\hat{J}$, i.e., we argue that
\begin{equation}\label{eq:Ksub}
	\langle\Psi_{\lambda}^{\rm HF}|\hat{K}|\Psi_{\lambda}^{\rm HF}\rangle = O(\lambda^{-1/2})\qquad(\lambda\to\infty). 
\end{equation}
Before we shall support this conjecture with a variational argument, we discuss its consequences. 

If Eq.~\eqref{eq:Ksub} holds, then $\Psi_{\lambda}^{\rm HF}$ for $\lambda\to\infty$ ends up minimizing the even simpler operator $\lambda(\hat{V}_{ee}-\hat{J})$,
\begin{eqnarray}\label{eq:PsiHFlambdainf0}
	\lim_{\lambda\to\infty}\Psi_\lambda^{\rm HF} & = & \mathop{\rm argmin}_\Psi\langle\Psi|\hat{V}_{ee}-\hat{J}|\Psi\rangle,\\
	\lim_{\lambda\to\infty}   W_\lambda^{\rm HF} & = & \min_\Psi\langle\Psi|\hat{V}_{ee}-\hat{J}|\Psi\rangle+\nonumber\\
                                                 &   & +\,U[\rho^{\rm HF}]+2E_x^{\rm HF}+O(\lambda^{-1/2})
	\label{eq:WHFlambdainf0}
\end{eqnarray}
The ``asymptotic hamiltonian'' $\hat{\cal H}_\infty^{\rm HF}=\hat{V}_{ee}-\hat{J}[\rho^{\rm HF}]$ is completely specified by the HF density $\rho^{\rm HF}(\rv)$, since $N=\int d\rv\,\rho^{\rm HF}(\rv)$ and $\hat{J}[\rho]=\sum_{i=1}^N v_{\rm H}(\rv_i;[\rho])$, with $v_{\rm H}(\rv;[\rho])\equiv \int \frac{\rho(\rv')}{|\rv-\rv'|}\,d\rv'$.
Consequently, also the minimizer in Eqs.~\eqref{eq:PsiHFlambdainf0} and \eqref{eq:WHFlambdainf0} is specified solely by $\rho^{\rm HF}$,
\begin{equation} 
	\label{eq:PsiHFlambdainf}
	\lim_{\lambda\to\infty}\Psi_\lambda^{\rm HF} = \Psi_\infty^{\rm HF}[\rho^{\rm HF}],
\end{equation}
and the minimum in Eq.~\eqref{eq:WHFlambdainf0} is a functional of $\rho^{\rm HF}$,
\begin{eqnarray}
		\lim_{\lambda\to\infty}   W_\lambda^{\rm HF} & = & E_{\rm el}[\rho^{\rm HF}]+2E_x^{\rm HF}+O(\lambda^{-1/2}).
		\label{eq:lambdainfHFall}
\end{eqnarray}
The minimizer in Eq.~\eqref{eq:PsiHFlambdainf0} could be not unique, but this does not affect the value of the minimum, which is the object of the present investigation.
The functional $E_{\rm el}[\rho]=\min\limits_\Psi\langle\Psi|\hat{V}_{ee}-\hat{J}[\rho]|\Psi\rangle+U[\rho]$ has a simple classical interpretation: Since $\hat{\cal H}_\infty^{\rm HF}=\hat{V}_{ee}-\hat{J}[\rho^{\rm HF}]$ is a purely multiplicative operator, 
\begin{equation}
	\hat{\cal H}_\infty^{\rm HF}=\sum_{\substack{i,j=1 \\ j>i}}^N\frac{1}{|\rv_i-\rv_j|}-\sum_{i=1}^N v_{\rm H}(\rv_i;[\rho^{\rm HF}]),
\label{eq:HinftyHF}
\end{equation}
the square modulus $|\Psi_\infty^{\rm HF}|^2$ of its minimizing wave function is a distribution in ${\mathbb{R}}^{3N}$ that is zero wherever 
$\hat{\cal H}_\infty^{\rm HF}$ as a function of $\rv_1,...,\rv_N$ does not assume its global minimum (if it were otherwise it would not be optimal as we could always lower the energy by increasing the weight of the wave function in the global minimum of $\hat{\cal H}_\infty^{\rm HF}$ ). In other words,
\begin{equation}\label{eq:EelHF}
	E_{\rm el}[\rho]\equiv \min_{\{\rv_1\dots\rv_N\}}\left\{\sum_{\substack{i,j=1 \\ j>i}}^{N}\frac{1}{|\rv_i-\rv_j|}-\sum_{i=1}^N v_{\rm H}(\rv_i;[\rho])+U[\rho]\right\}
\end{equation}
is the minimum total electrostatic energy of $N$ equal classical point charges $(-e)$ in a positive background with continuous charge density $(+e)\rho(\rv)$. The term $U[\rho]$, inherited from Eq.~\eqref{eq:WlambdaHF}, represents the background-background repulsion.
%\begin{equation}\label{eq:PsiinfHFexpl}
%	|\Psi_\infty^{\rm HF}(\rv_1,...,\rv_N)|^2={\cal N}\sum_{\alpha}\prod_{i=1}^N \delta(\rv_i-\rv_i^{\rm min}(\alpha)).
%\end{equation}

Strictly speaking, the minimizer $\Psi_\infty^{\rm HF}$ is not in the space of allowed wavefunctions, so that the minimum is actually an infimum, similarly to the DFT case.\cite{CotFriKlu-CPAM-13,CotFriKlu-ARMA-18,Lew-CRM-18}

Equations \eqref{eq:PsiHFlambdainf}-\eqref{eq:lambdainfHFall} comprise a central result of this work: they show that the strong-interaction limit of the HF adiabatic connection can be determined from a functional of the HF density, providing some theoretical justification for resumming the MP series by using a DFT-like expansion at large $\lambda$ with functionals of $\rho^{\rm HF}$,\cite{FabGorSeiDel-JCTC-16,GiaGorDelFab-JCP-18,VucGorDelFab-JPCL-18} although, as we will discuss, there are still several points to be addressed.

We now analyze the functionals $\Psi_\infty^{\rm HF}[\rho]$ and $E_{el}[\rho]$, comparing them with the DFT case.  As $\lambda\to\infty$, $\hat{H}_\lambda^{\rm DFT}$ of Eq.~\eqref{eq:adiabDFT} tends to $\lambda\,\hat{\cal H}_\infty^{\rm DFT}[\rho]$, with\cite{SeiGorSav-PRA-07} 
\begin{equation}\label{eq:HinftyDFT}
	\hat{\cal H}_\infty^{\rm DFT}[\rho]=\sum_{\substack{i,j=1 \\ j>i}}^N\frac{1}{|\rv_i-\rv_j|}+\sum_{i=1}^N v_\infty(\rv_i,[\rho]).
\end{equation}
Comparing $\hat{\cal H}_\infty^{\rm DFT}[\rho]$ with $\hat{\cal H}_\infty^{\rm HF}[\rho]$ of Eq.~\eqref{eq:HinftyHF}, we see that both hamiltonians consist of the electron-electron repulsion operator and of an attractive one-body potential. In the HF case the attractive potential is $-v_{\rm H}(\rv,[\rho])$, which is, for typical Hartree-Fock densities, strong enough to create a classical bound crystal. To be more precise, $-v_{\rm H}(\rv,[\rho])$ is more attractive than the one-body potential $v_\infty(\rv,[\rho])$. In fact, the potential $v_\infty(\rv,[\rho])$, which has been studied in several works\cite{Sei-PRA-99,SeiGorSav-PRA-07,VucLevGor-JCP-17,GiaVucGor-JCTC-18} is generated by a charge that integrates to $N-1$, \cite{SeiGorSav-PRA-07,VucLevGor-JCP-17}
\begin{equation}
	\frac{1}{4\pi}\int \nabla^2 v_\infty(\rv,[\rho]) \,d\,\rv= N-1,
\end{equation}
while the attractive potential $-v_{\rm H}(\rv,[\rho])$ is generated by the given  density $\rho(\rv)$, which integrates to $N$. For finite systems, the state $\Psi_\infty^{\rm HF}[\rho]$ is thus more compact than the state $\Psi_\infty^{\rm DFT}[\rho]$: this is due to the density constraint in the DFT adiabatic connection, which forces $\Psi_\infty^{\rm DFT}[\rho]$ to have the given quantum mechanical density $\rho(\rv)$.\cite{SeiGorSav-PRA-07,GiaVucGor-JCTC-18}

We note in passing that, for given occupied HF orbitals, we have the chain of inequalities
\begin{equation}
	W_\infty^{\rm HF} \le E_{el}[\rho^{\rm HF}]\le W_\infty^{\rm DFT}[\rho^{\rm HF}].
\end{equation}
The first one, $W_\infty^{\rm HF} \le E_{el}[\rho^{\rm HF}]$, is trivial since $W_\infty^{\rm HF}=E_{el}[\rho^{\rm HF}]+2 \,E_x^{\rm HF}$ and $E_x^{\rm HF}\le 0$. The second inequality holds for any density $\rho(\rv)$, $E_{el}[\rho]\le W_\infty^{\rm DFT}[\rho]$. To prove it, we introduce the bifunctional $\mathcal{W}[\rho,v]$,
\begin{equation}
	\mathcal{W}[\rho,v]=\inf_\Psi\langle\Psi|\hat{V}_{ee}-\sum_{i=1}^N v(\rv_i)|\Psi\rangle+\int\rho(\rv)v(\rv)\,d\rv,
\end{equation}
for which we have
\begin{equation} \label{eq:bifHF}
	\mathcal{W}[\rho,v_{\rm H}[\rho]]=E_{el}[\rho]+U[\rho],
\end{equation}
and, from the dual formulation of $W_\infty^{\rm DFT}[\rho]$,\cite{SeiGorSav-PRA-07,ButDepGor-PRA-12,VucWagMirGor-JCTC-15}
\begin{equation} \label{eq:dualDFT}
	W_\infty^{\rm DFT}[\rho]+U[\rho]=\max_v \mathcal{W}[\rho,v],
\end{equation}
which clearly completes the proof. 

As promised, we provide a variational argument to support the assumption of Eq.~\eqref{eq:Ksub}, sketching the main points and leaving a more detailed treatment to a longer paper. We start by considering the global minimum ${ \underline R^{\rm min}}\equiv\{\rv_1^{\rm min},\dots,\rv_N^{\rm min}\}$ of the function $\hat{\cal H}_\infty^{\rm HF}$ of Eq.~\eqref{eq:HinftyHF}, and construct the simple trial wavefunction
\begin{equation}
	\Psi_\lambda^{T}(\rv_1,\dots,\rv_N)=\prod_{i=1}^N G_{\alpha(\lambda)}(\rv_i-\rv_i^{\rm min}),
	\label{eq:PsiT}
\end{equation}
where $G_\alpha(\rv)=\frac{\alpha^{3/4}}{\pi^{3/4}}e^{-\frac{\alpha}{2}|\rv|^2}$,
with $\alpha$ a $\lambda$-dependent variational parameter that goes to infinity for large $\lambda$,  $\alpha(\lambda)\sim \lambda^q$ with $q>0$. By construction, when $\alpha\to\infty$ (i.e., when $\lambda\to\infty$) we  have that
\begin{equation}
	\lim_{\lambda\to\infty}|\Psi_\lambda^T[\rho^{\rm HF}]|^2=|\Psi_\infty^{\rm HF}[\rho^{\rm HF}]|^2
\end{equation}
where $\Psi_\infty^{\rm HF}$ was introduced in Eq.~\eqref{eq:PsiHFlambdainf0}  (in the case of degeneracy we can select one of the minimizers, since here we only want to obtain an upper bound to the lowest eigenvalue of $H_\lambda^{\rm HF}$). We now analyze, for large $\alpha$, the expectation value on $\Psi_\lambda^{T}$ of each term appearing in $\hat{H}_\lambda^{\rm HF}$ of Eq.~\eqref{eq:HlambdaHF}, obtaining
\begin{align}\label{eq:TlambdainfHF}
&	\langle\Psi_\lambda^T|\hat{T}|\Psi_\lambda^T\rangle = t\,\alpha \\
&	\langle\Psi_\lambda^T|\lambda(\hat{V}_{ee}-\hat{J})|\Psi_\lambda^T\rangle  = \lambda(E_{el}-U)+\lambda\left(\frac{h}{\alpha}+o(\alpha^{-1})\right) \label{eq:NextTermVee-J} \\
&	\langle\Psi_\lambda^T|-\lambda\,\hat{K}|\Psi_\lambda^T\rangle  = \lambda\left(\frac{k}{\alpha}+o(\alpha^{-1})\right)\label{eq:Wk} \\
& \langle\Psi_\lambda^T|\hat{V}_{\rm ext}+\hat{J}+\hat{K}|\Psi_\lambda^T\rangle \sim O(\alpha^0), \label{eq:order0}
\end{align}
where $t$, $h$, and $k$ are all {\em positive} numbers. This is obvious for $t$, but it is also true for $k$ because the expectation of $-\hat{K}$ is positive for any wavefunction $\Psi$, as $\hat{K}$ has a negatively definite kernel. The fact that the expectation value of $\hat{K}$ on $\Psi_\lambda^{T}$ vanishes as $\alpha^{-1}$ for large $\alpha$ is due to the non-locality of $\hat{K}$, which samples the gaussians in the bra and in the ket in different points of space, and to the regularity properties of the HF orbitals (which have no delta-function singularities). The positivity of $h$ in Eq.~\eqref{eq:NextTermVee-J} can be proven by expanding $\hat{\cal H}_\infty^{\rm HF}$ around ${ \underline R^{\rm min}}$ up to second order, which gives an hessian matrix positive definite. 

Putting together Eqs.~\eqref{eq:TlambdainfHF}-\eqref{eq:order0} and replacing $\alpha$ with $\lambda^q$ we find that, for large $\lambda$, the expectation value of $\hat{H}_\lambda^{\rm HF}$ on $\Psi_\lambda^T$ behaves asymptotically as
\begin{equation}
	\langle\Psi_\lambda^T| \hat{H}_\lambda^{\rm HF}|\Psi_\lambda^T\rangle= \lambda(E_{el}-U)+ t \lambda^q+(h+k)\lambda^{1-q}+o(\lambda^{1-q}).
\end{equation}
Being $t$, $h$ and $k$ positive, we see that the best variational choice to make the next leading term after $O(\lambda)$ increase  with the lowest possible power of $\lambda$ is $q=1/2$, as conjectured in Eq.~\eqref{eq:Ksub}. Although $\Psi_\lambda^T$ of Eq.~\eqref{eq:PsiT} is not antisymmetric, we can always properly antisymmetrize it, which only leads to corrections $O(e^{-\alpha})$ in the computation of the expectation values, similarly to the DFT case.\cite{GroKooGieSeiCohMorGor-JCTC-17}

Thus, we have explicitly constructed a variational wavefunction that yields the minimum possibile value for the leading term $O(\lambda)$ in the expectation of $\hat{H}_\lambda^{\rm HF}$. In fact, since $E_{el}[\rho^{\rm HF}]-U[\rho^{\rm HF}]$ is the global minimum of the multiplicative operator $\hat{V}_{ee}-\hat{J}$, there is no wavefunction that can yield a lower expectation for this operator. Moreover, since $-\hat{K}$ is positive definite, the best we can do is to make its expectation zero when $\lambda\to\infty$, which our wavefunction is able to do.

This variational argument also shows that the next leading term in $W_\lambda^{\rm HF}$ should be order $\lambda^{-1/2}$, similarly to the DFT case of Eq.~\eqref{eq:lambdainfDFT}. A quantitative estimate of this next leading term could be in principle obtained by using the normal modes around the minimum of $\hat{V}_{ee}-\hat{J}$:   a unitary transformation from the ${\bf r}_i-\rv_i^{\rm min}$ to the normal modes coordinates $ \xi_1,\dots,\xi_{3N}$ that diagonalize the hessian of $\hat{\cal H}_\infty^{\rm HF}$ at ${ \underline R^{\rm min}}$ leads to a set of uncoupled harmonic oscillators whose spring constant scales with $\lambda$,
\begin{equation}
	\hat{H}_\lambda^{\rm ZP}=-\frac{1}{2}\sum_{\alpha=1}^{3N} \frac{\partial^2}{\partial \xi_\alpha^2}+\frac{\lambda}{2}\sum_{\alpha=1}^{3N}\omega_\alpha^2 \xi_\alpha^2,
\end{equation}
with $\omega_\alpha^2$ the eigenvalues of the hessian of $\hat{\cal H}_\infty^{\rm HF}$ at ${ \underline R^{\rm min}}$. The ground-state of $\hat{H}_\lambda^{\rm ZP}$ is obtained by occupying the lowest state of each oscillator, with the product state 
\begin{equation}
	\Psi_\lambda^{\rm ZP}(\xi_1,\dots\,\xi_{3N})=\prod_{\alpha=1}^{3N} \frac{(\omega_\alpha\sqrt{\lambda})^{1/4}}{\pi^{1/4}} e^{-\sqrt{\lambda}\,\omega_\alpha\frac{\xi_\alpha^2}{2}}.
\end{equation}
This wavefunction should provide the minimum possible expectation, to order $\lambda^{1/2}$, of $\hat{T}+\lambda(\hat{V}_{ee}-\hat{J})$. However, since $-\lambda \hat{K}$ is of the same order $\lambda^{1/2}$, we cannot exclude at this point that the minimization of the full $\hat{T}+\lambda(\hat{V}_{ee}-\hat{J}-\hat{K})$ could lead to a different set of occupied oscillator states. This investigation will be the object of future works.
From our present treatment we have so far
\begin{equation}
	W_{\lambda\to\infty}^{\rm HF} = W_\infty^{\rm HF}+\frac{1}{\sqrt{\lambda}}W_\infty'^{\rm HF}+\dots,
\end{equation}
with
\begin{align}
	 & W_\infty^{\rm HF}=E_{el}[\rho^{\rm HF}]+2\, E_x^{\rm HF} \label{eq:WinftyHFfinal}\\
	 & W_\infty'^{\rm HF}=\frac{1}{2}\sum_{\alpha=1}^{3N}\frac{\omega_\alpha[\rho^{\rm HF}]}{2}+W_{K,\infty}'^{\rm HF},
	 \label{eq:WZPHFfinal}
\end{align}
where $W_{K,\infty}'^{\rm HF}$ is due to the effect of $-\lambda\hat{K}$ at orders $\lambda^{1/2}$ in $\hat{H}_\lambda^{\rm HF}$ and is a functional of the occupied HF orbitals. Eq.~\eqref{eq:WinftyHFfinal} should be exact while Eq.~\eqref{eq:WZPHFfinal} is for now a conjecture. We also see that both $W_\infty^{\rm HF}$ and $W_\infty'^{\rm HF}$ have a part that is a functional of the HF density only, and a part that is a functional of the occupied HF orbitals. In both cases, the part that is a density functional has an origin similar to the one of the DFT functionals of Eq.~\eqref{eq:lambdainfDFT}, being, respectively, a classical electrostatic energy and the potential energy of zero-point oscillations around a classical minimum. The parts that need the knowledge of the occupied HF orbitals do not appear in the DFT case. This structure should be exact, although the detailed form of $W_\infty'^{\rm HF}$ might include a different set of occupied oscillator states.

Although the $\lambda\to\infty$ limit of $W_\lambda^{\rm HF}$ has a structure similar to the one of DFT, there are many differences that need to be kept in mind. Both $W_\lambda^{\rm DFT}[\rho]$ and $W_\lambda^{\rm HF}$ are decreasing functions of $\lambda$,
\begin{equation}
	\frac{d}{d\lambda} W_\lambda^{\rm DFT}[\rho]\le 0,\qquad \frac{d}{d\lambda} W_\lambda^{\rm HF}\le 0,
\end{equation}
but $W_\lambda^{\rm DFT}[\rho]$  for $\lambda\ge 0$ is believed to be convex or at least piecewise convex (if there are crossings of states), while $W_\lambda^{\rm HF}$ is for sure not always convex. In fact, the MP2 correlation energy usually underestimates (in absolute value) the total correlation energy $E_c^{\rm HF}$, implying that $W_\lambda^{\rm HF}$ for $0<\lambda\ll 1$ must run below its tangent; thus, $W_\lambda^{\rm HF}$ usually starts concave for small $\lambda$ and then needs to change convexity to tend to the finite asymptotic value $W_\infty^{\rm HF}$ for large $\lambda$. Moreover, while the density constraint of the DFT adiabatic connection usually mitigates the crossing of states, the HF adiabatic connection might have jumps or kinks as $\lambda$ is increased. A simple example is the $N=1$ case, for which $W_\lambda^{\rm HF}=-U[\rho^{\rm HF}]$ for $0\le\lambda\le 1$, while for $\lambda>1$ the curve starts to decrease, tending, as $\lambda\to\infty$ to a well defined value, with the electrostatic energy determined by the configuration in which the electron is sitting in the minimum of $-v_{\rm H}(\rv,[\rho])$. 

In conclusion, we have shown that by looking at the $\lambda\to\infty$ limit of the HF adiabatic connection we recover functionals of the HF density, revealing a new intriguing formal link between HF and DFT. However, we should also stress that
the use of models for $W_\lambda^{\rm HF}$ taken from DFT, although somehow justified by our analysis, should at this stage still be taken with some caution. The empirical observation so far,\cite{GiaGorDelFab-JCP-18,VucGorDelFab-JPCL-18} is that these models are not accurate for total energies, but work rather well for interaction energies, with a small variance, particularly for non-covalent complexes.\cite{GiaGorDelFab-JCP-18,VucGorDelFab-JPCL-18} This point requires further investigation, which will be the object of a paper in preparation, where $W_\lambda^{\rm HF}$ will be computed and analyzed for various systems, and will be compared against various models. We will also evaluate and further analyze $W_\infty^{\rm HF}$ and $W_\infty'^{\rm HF}$, making a detailed numerical comparison with the corresponding DFT functionals. We can already remark that the difference between $W_\infty^{\rm HF}$ and $W_\infty^{\rm DFT}$ can be big. For example, for the He atom we have $W_\infty^{\rm HF}\approx-4.347$~Ha, while $W_\infty^{\rm DFT}\approx-1.50$~Ha.
Other promising research lines opened by this study is to investigate whether it is possible to extract a model for the self-energy in the strong-coupling limit, to be used in the context of Green's functions approaches,\cite{Hed-PR-65,SchAng-JCP-89,TarRomBerRei-PRB-17,LosRomBer-JCTC-18} and to analyze in the same spirit adiabatic connections appearing in other theories.\cite{Per-PRL-18,Per-IJQC-18}

This work was supported by the European Research Council under H2020/ERC Consolidator Grant corr-DFT (Grant No. 648932). S.V. acknowledges financial support from NWO through Rubicon grant 019.181EN.026. 

%merlin.mbs aipnum4-1.bst 2010-07-25 4.21a (PWD, AO, DPC) hacked
%Control: key (0)
%Control: author (8) initials jnrlst
%Control: editor formatted (1) identically to author
%Control: production of article title (-1) disabled
%Control: page (0) single
%Control: year (1) truncated
%Control: production of eprint (0) enabled

%\bibliography{bib_clean}
%merlin.mbs aipnum4-1.bst 2010-07-25 4.21a (PWD, AO, DPC) hacked
%Control: key (0)
%Control: author (8) initials jnrlst
%Control: editor formatted (1) identically to author
%Control: production of article title (-1) disabled
%Control: page (0) single
%Control: year (1) truncated
%Control: production of eprint (0) enabled
%

\end{document}